\documentclass[fleqn,usenatbib]{mnras}

\usepackage{newtxtext,newtxmath}

\usepackage[T1]{fontenc}

\DeclareRobustCommand{\VAN}[3]{#2}
\let\VANthebibliography\thebibliography
\def\thebibliography{\DeclareRobustCommand{\VAN}[3]{##3}\VANthebibliography}

\usepackage{graphicx}	
\usepackage{amsmath}	

\usepackage[switch]{lineno}
\usepackage{hyperref}
\usepackage{color}
\usepackage{multirow}
\usepackage{xspace}
\usepackage{soul}

\newcommand{\ergflux}{\ensuremath{\mathrm{erg\,cm^{-2}\,s^{-1}}}}
\newcommand{\gm}{\ensuremath{\gamma}}
\newcommand{\fermi}{\textit{Fermi}~}
\newcommand{\swift}{\textit{Swift}~}

\newcommand{\xmm}{\textit{XMM-Newton}\xspace}

\title[Mrk 421: Rapid Variability During Extreme Flaring]{Rapid Variability of Mrk 421 During Extreme Flaring as Seen Through the Eyes of \xmm}

\author[Gokus et al.]{%
A.~Gokus,$^{1,2,3}$\thanks{E-mail: gokus@wustl.edu, dr.andrea.gokus@gmail.com}
J.~Wilms,$^{2}$
M.~Kadler,$^{3}$
D.~Dorner,$^{4,3,\dagger}$
M.~A.~Nowak,$^1$
A.~Kreikenbohm,$^{3}$
K.~Leiter,$^{3}$
\newauthor
T.~Bretz,$^{4,5,\dagger}$
B.~Schleicher,$^{4,3,\dagger}$
A.~G.~Markowitz,$^{6,7}$
K.~Pottschmidt,$^{8,9}$
K.~Mannheim,$^{3,\dagger}$
\newauthor
I.~Kreykenbohm,$^2$
M.~Langejahn,$^{3,10}$
F.~McBride,$^{11}$
T.~Beuchert,$^{2}$
T.~Dauser,$^2$
M.~Kreter,$^{12}$
\newauthor
\textit{and the FACT Collaboration}:
J.~Abhir,$^{4}$
D.~Baack,$^{13}$
M.~Balbo,$^{14}$
A.~Biland,$^{4}$
K.~Brand,$^{3}$
J.~Buss,$^{13}$
\newauthor
L.~Eisenberger,$^{3}$
D.~Elsaesser,$^{13}$
P.~Günther,$^{3}$
D.~Hildebrand,$^{4}$
M.~Linhoff,$^{13}$
A.~Paravac,$^{3}$
\newauthor
W.~Rhode,$^{13}$
V.~Sliusar,$^{14}$
S.~Hasan,$^{4}$
R.~Walter$^{14}$
\\
$^{1}$Department of Physics \& McDonnell Center for the Space Sciences, Washington University in St. Louis, One Brookings Drive, St. Louis, MO 63130, USA\\
$^{2}$Dr.\ Karl Remeis-Sternwarte \& ECAP, Universit\"at Erlangen-N\"urnberg, Sternwartstr. 7, 96049 Bamberg, Germany\\
$^{3}$Institut f\"ur Theoretische Physik und Astrophysik, Universit\"at W\"urzburg, Emil-Fischer-Straße 31, 97074 W\"urzburg, Germany\\
$^{4}$ETH Z\"urich, Institute for Particle Physics and Astrophysics, Otto-Stern-Weg 5, 8093 Z\"urich, Switzerland\\
$^5$also at GSI Helmholtzzentrum für Schwerionenforschung GmbH, Darmstadt, Germany\\
$^6$Nicolaus Copernicus Astronomical Center, Polish Academy of Sciences, ul. Bartycka 18, 00-716, Warszawa, Poland \\
$^7$University of California, San Diego, Center for Astrophysics and Space Sciences, MC 0424, La Jolla, CA, 92093-0424, USA \\
$^8$CRESST and CSST, University of Maryland, Baltimore County, 1000 Hilltop Circle, Baltimore, MD 21250, USA \\
$^9$Code 661 Astroparticle Physics Laboratory, NASA Goddard Space Flight Center, Greenbelt, MD 20771, USA \\
$^{10}$Lehrstuhl f\"ur Data Science in Earth Observation, Technische Universit\"at M\"unchen , Arcisstra{\ss}e 21, 80333 M\"unchen, Germany \\
$^{11}$Department of Physics and Astronomy, Bowdoin College, Brunswick, ME 04011, USA \\
$^{12}$Centre for Space Research, North-West University, Potchefstroom 2520, South Africa \\
$^{13}$TU Dortmund, Experimental Physics 5, Otto-Hahn-Str.\ 4a, 44227 Dortmund, Germany \\
$^{14}$University of Geneva, Department of Astronomy, Chemin d'Ecogia 16, 1290 Versoix, Switzerland \\
$^{\dagger}$also member of the FACT Collaboration
}

\date{Accepted XXX. Received YYY; in original form ZZZ}

\pubyear{2024}

\begin{document}
\label{firstpage}
\pagerange{\pageref{firstpage}--\pageref{lastpage}}
\maketitle

\begin{abstract}
By studying the variability of blazars across the electromagnetic spectrum, it is possible to resolve the underlying processes responsible for rapid flux increases, so-called flares.
We report on an extremely bright X-ray flare in the high-peaked BL Lacertae object Mrk\,421 that occurred simultaneously with enhanced \gm-ray activity detected at very high energies (VHE) by FACT on 2019 June 9. We triggered an observation with \xmm, which observed the source quasi-continuously for 25\,hours. We find that the source was in the brightest state ever observed using \xmm, reaching a flux of $2.8\times10^{-9}$ \ergflux over an energy range of 0.3\,--\,10\,keV. We perform a spectral and timing analysis to reveal the mechanisms of particle acceleration and to search for the shortest source-intrinsic time scales.
Mrk\,421 exhibits the typical harder-when-brighter behaviour throughout the observation and shows a clock-wise hysteresis pattern, which indicates that the cooling dominates over the acceleration process. While the X-ray emission in different sub-bands is highly correlated, we can exclude large time lags as the computed zDCFs are consistent with a zero lag.
We find rapid variability on time scales of 1\,ks for the 0.3\,--\,10\,keV band and down to 300\,s in the hard X-ray band (4\,--\,10\,keV). Taking these time scales into account, we discuss different models to explain the observed X-ray flare, and find that a plasmoid-dominated magnetic reconnection process is able to describe our observation best.
\end{abstract}




\section{Introduction}
Among active galactic nuclei (AGN), blazars show the strongest and the most rapid variability \citep[e.g.,][]{stein1976,wagner1995,ulrich1997}.
Their luminosity and extreme properties originate from a collimated outflow, the so-called jet, of relativistically moving particles close to the line of sight \citep{antonucci1993,urry1995}, and their emission is Doppler boosted towards the observer.
Their characteristic broadband spectral energy distribution has a double-hump structure and sub-classes of blazars are broadly defined by their overall luminosity and the peak position of the first hump, which lies somewhere between near-IR to X-ray energies. The emission of the low-energy hump is typically strongly dominated by leptonic synchrotron radiation, while the high-energy hump, which peaks at \gm-ray energies, can be explained via leptonic Inverse Compton emission \citep[e.g.,][]{ghisellini1985,maraschi1992} as well as hadronic processes \citep[e.g.,][]{mannheim1993,atoyan2001,mucke2001,mucke2003,kelner2008,bottcher2013}.

Blazar variability is observed across the entire electromagnetic spectrum and from sub-hour time scales to years \citep[e.g.,][]{stein1976,wagner1995,ulrich1997,pian2002,bhatta2020}. Many coordinated multi-wavelength studies, partially with a focus on high-amplitude flux increases, that is flares, have been conducted to shed light on the emission regions within the jet as well as the underlying particle acceleration processes.
The fastest variability is typically observed during bright flares and can happen on sub-hour time scales. The most rapid flux variations, on timescales below 10 minutes, have been observed at \gm-ray energies for five blazars so far: PKS\,2155$-$304 \citep{aharonian2007}, Mrk\,501 \citep{albert2007}, IC\,301 \citep{aleksic2014}, 3C\,279 \citep{ackermann2016}, and CTA\,102 \citep{meyer2019}.
These fluctuations challenge established models such as the shock-in-jet model \citep{marscher1985} or the spine-sheath-model \citep{henri1991,ghisellini2005}, and motivated the development of new scenarios, such as the minijet-in-a-jet model \citep[e.g.,][]{ghisellini2008, giannios2009,narayan2012,zhang2021}, with slightly different underlying processes, or the jet interacting with a star or a gas cloud \citep{barkov2012,heil2020}. Alternatively, stochastic perturbations in the particle acceleration process are able to explain the observed, log-normal flux distributions, which are characteristic for blazars \citep[e.g.,][]{sinha2018,burd2021}.

Due to its proximity \citep[$z=0.031$;][]{ulrich1975} and brightness, which enables obtaining high signal-to-noise data at all wavelengths, the high-synchrotron peaked (HSP) source Markarian 421 (Mrk\,421) has been extensively monitored. 
In 1992, it was detected as an extragalactic TeV-emitter with the Whipple telescope \citep{punch1992}, and since then has exhibited frequent and bright \gm-ray flares with the flux exceeding that of the Crab nebula at TeV energies by a factor of three or more \citep[e.g.,][]{albert2007_mrk421,bartoli2016}, even over ten Crab flux units in three instances \citep{acciari2014}.
While a large amount of data has been gathered during different activity states of the source, as well as transitioning states, the behaviour of Mrk\,421 has not been fully understood because of its complexity.
The source shows signatures of rapid variability, i.e., on sub-hour time scales at VHE \gm-ray energies \citep[e.g.,][]{gaidos1996,abeysekara2020,acciari2020, magic2021} as well as at X-ray energies \citep[e.g.,][]{abeysekara2017,chatterjee2021,noel2022}.
Its TeV \gm-ray and X-ray emission have been found to be usually correlated during flares \citep[e.g.,][]{macomb1995,buckley1996,albert2007_mrk421,fossati2008,acciari2011,cao2013}, but also during quiescent phases \citep{horan2009,aleksic2015,paneque2021}.
However, Mrk\,421 has also shown indications for so-called orphan flares at VHE \gm-rays \citep{blazejowski2005,acciari2011, fraija2015}, which lack correlated X-ray flares and whose origins are unclear.
In a study covering a time span of 5.5 years and densely sampled data in the radio, X-ray, and \gm-ray (GeV and TeV) energy range, \citet{sliusar2019} report 29 flares that occurred simultaneously in the X-ray and TeV band, and two orphan TeV flares. On average, this would result in a \gm-ray flare occurring roughly every 65 days.
A 14-year long study with the Whipple telescope, covering a time span from 1995 to 2009, determined an average flux of $0.446\pm0.008$ Crab flux units for Mrk\,421 \citep{acciari2014}.

\begin{figure}
    \centering
    \includegraphics[width=\columnwidth]{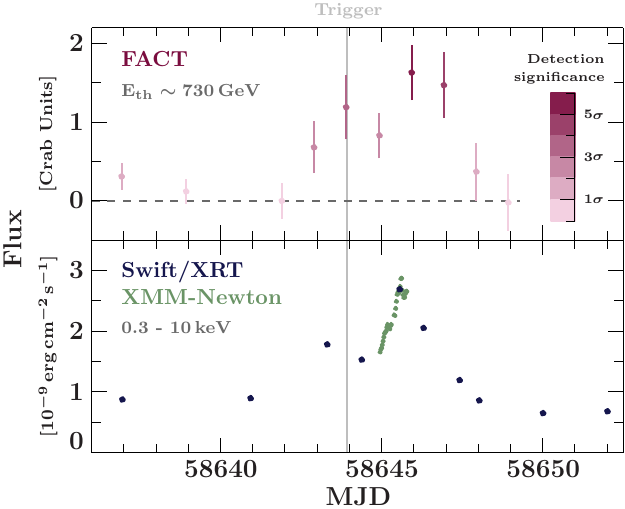}
    \caption{VHE \gm-ray (top) and X-ray (bottom) light curve covering the time span from 2019 June 2 to 2019 June 18. Data of Mrk\,421 at \gm-ray energies (E$_{\mathrm{th}}\sim$730\,GeV) was obtained through the long-term monitoring campaign by FACT, while monitoring of the source in the X-ray band (0.3--10\,keV) was done with \swift-XRT (dark blue), which is taken from \citet{gokus2021_mrk421}. The \xmm data (teal) is included as well and covers the majority of the X-ray flare. The color scale for the \gm-ray light curve bins displays the detection significance for Mrk\,421 for each observation. The inlay of the color scale does not cover any \gm-ray data since there was no visibility of Mrk\,421 after MJD\,58649 for FACT. Note that the uncertainties for the X-ray flux are so well constrained that error bars are not visible.}
    \label{fig:fact_xray_lc}
\end{figure}

In this work, we present an X-ray observation of Mrk\,421 that was performed with \xmm \citep{jansen2001}, and  coincident with high VHE \gm-ray activity in June 2019. In order to obtain such a simultaneous observation, we use the long-term monitoring campaign of nightly observations with the First G-APD Cherenkov Telescope \citep[FACT;][]{fact_instr} for selected blazars \citep{dorner2021,arbet-engels2021}.
Our campaign, which has been designed to trigger several multi-wavelength instruments during a VHE \gm-ray flare is described in \citet{kreikenbohm_phdthesis}, and the details on the follow-up observations during June 2019 are reported in \citet{gokus2021_mrk421} and \citet{gokus_phdthesis}. A light curve displaying the monitoring of Mrk\,421 in the \gm-ray and X-ray band in June 2019, the time of the extreme X-ray flare presented in this work, is shown in Fig.~\ref{fig:fact_xray_lc}.
The goal was to trigger multi-wavelength follow-up observations for the occurrence of a VHE \gm-ray flux above 2 Crab Units (CU). This condition for sending the requests for target-of-opportunity observations was fulfilled on 2019 June 9 based on the quick-look analysis, which is described in \citet{dorner2015}. However, this analysis does not include a correction for the dependence of the threshold on the zenith distance and ambient light \citep[see][for details]{arbet-engels2021}, and a subsequent analysis found that the \gm-ray flux (E$_{\mathrm{th}}\sim$730\,GeV) was slightly lower. 
The data were analysed with the Modular Analysis and Reconstruction Software \citep[MARS;][]{mars_software}, using the background suppression method described in \citet{beck2019}. Good quality data were selected based on the cosmic-ray rate \citep{beck2019,hildebrand2017} calculating the artificial trigger rate R750 above a threshold of 750~DAC\footnote{Digital-to-analog converter}-counts. A corrected rate, $R750_{\mathrm{cor}}$, is calculated for an evolving zenith distance \citep{mahlke2017,bretz2019}. Because the cosmic-ray rate changes from season to season, a monthly reference value $R750_{\mathrm{ref}}$ was derived, and a cut for good quality data for $0.93 < R750_{\mathrm{cor}}/R750_{\mathrm{ref}} < 1.3$ was applied.
The highest VHE \gm-ray flux is measured on MJD\,58645.9, which is coincident with the X-ray flare, and our analysis yields $1.63\pm0.35$\,CU, or $(5.1\pm1.1)\times10^{-11}$\,ph\,cm$^{-2}$\,s$^{-1}$.
Since the VHE \gm-ray flux did not quite reach the trigger threshold of 2\,CU (see Fig.~\ref{fig:fact_xray_lc}), we do not refer to the \gm-ray activity as a flare, but as high VHE \gm-ray activity.

Here, we focus on the variability analysis of the X-ray data obtained simultaneously with the \gm-ray activity.
With a peak 0.3--3\,keV flux of $1.6\times 10^{-9}\,\mathrm{erg}\,\mathrm{cm}^{-2}\,\mathrm{s}^{-1}$, the flux during our observation was the highest that has ever been detected with \xmm, and is close to the brightest 0.3--3\,keV fluxes ever measured for this source during a flare in 2013 April \citep[$2.2\times 10^{-9}\,\mathrm{erg}\,\mathrm{cm}^{-2}\,\mathrm{s}^{-1}$;][]{acciari2020} and a flaring episode in 2018 January \citep[$\sim2.8\times 10^{-9}\,\mathrm{erg}\,\mathrm{cm}^{-2}\,\mathrm{s}^{-1}$, estimated from the reported flux of $5\times 10^{-9}\,\mathrm{erg}\,\mathrm{cm}^{-2}\,\mathrm{s}^{-1}$ in the 0.3--10\,keV band;][]{kapanadze2020}\footnote{We note that this flux is extraordinary for an extragalactic object, as it is comparable to that of bright Galactic sources such as Cygnus X-1 \citep[e.g.,][]{grinberg2013,duro2016}.}.

This paper is structured as follows:
In Section~\ref{sec:xmmobs}, we describe the data extraction of the \xmm observation. We elaborate on the spectral analysis of these data in Section~\ref{sec:spectral_variability}, and perform a timing analysis in Section~\ref{sec:time_series_analysis}. We discuss the implications of our findings and conclude in Section~\ref{sec:discussion}.

\begin{figure*}
    \centering
    \includegraphics[width=\textwidth]{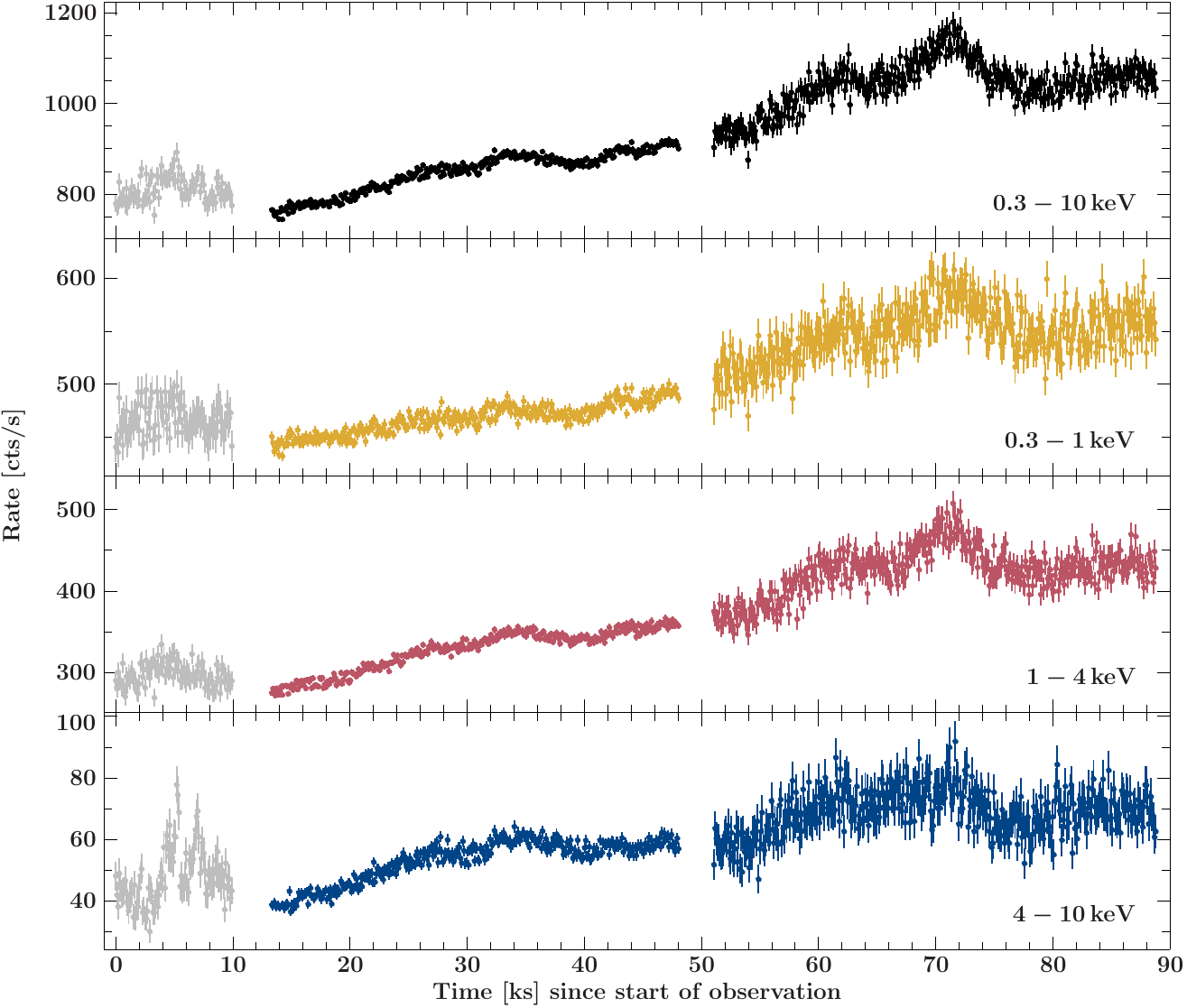}
    \caption{100s-binned light curve of Mrk\,421 taken with \xmm. The upper panel displays the full energy range (0.3--10\,keV), while the subsequent panels show the varying flux in the soft (0.3--1\,keV), medium (1--4\,keV) and hard (4--10\,keV) band, respectively. The data taken in the first 10\,ks (light gray) are not taken further into account due to a high risk of being contaminated by strong background flaring.}
    \label{fig:100sbinned_lcs_energydep}
\end{figure*}

\section{Observation with \xmm}\label{sec:xmmobs}
The \xmm observation presented in this work was performed between 2019 June 10 19:11:27 UTC and 2019 June 11 20:26:42 UTC (ObsID: 0845000901) as a target-of-opportunity observation after enhanced VHE \gm-ray activity was observed by FACT at 2019 June 9 22:30 UTC. We concentrate on the observations with the European Photon Imaging Camera (EPIC) pn-camera \citep{epic-pn}. Observations done with EPIC MOS-camera \citep{epic-mos} were heavily piled-up and could not be used for scientific analysis. Two changes of observing mode split the observation in three sections. The first 10\,ks were observed in the EPIC-pn Burst mode, followed by 35\,ks in Timing mode, and lastly 38\,ks in the Burst mode again. Between these sections, gaps of roughly 3\,ks length exist. The reason for the second switch of observation mode, i.e., from Timing to Burst mode, was the rise in flux by the source, which meant that the observation was no longer feasible in Timing mode.
We extract the EPIC-pn data following the standard methods of the \xmm Science Analysis System (SAS, Version 20.0). The part of the observation taken in Timing mode was affected by pile-up. We mitigate the pile-up effects by ignoring the central five CCD rows around the peak of the photon distribution. The extraction region for the Timing mode data is therefore \texttt{$20\leq$ RAWX $\leq35$ \&\& $41\leq$ RAWX $\leq56$}. The extraction region for the Burst mode data covers 40 rows,  \texttt{$20\leq$ RAWX $\leq60$}. For these data, we apply an additional cut at \texttt{RAWY}$\leq140$ in order to avoid pile-up that can occur during the readout \citep[see][]{kirsch2006}. 

For the full exposure time, we generate light curves with different time binnings (100\,s and 1\,ks) and for different energy ranges for the variability analysis. The 100\,s-binned light curve for both the full energy range (0.3--10\,keV) as well as the soft (0.3--1\,keV), medium (1--4\,keV) and hard (4--10\,keV) band are shown in Fig.~\ref{fig:100sbinned_lcs_energydep}. 
In addition, we extract a light curve with time binnings of 10\,ms for the data that were taken in the Timing mode, which are used for computing the power spectral density (see Sect.~\ref{sec:search_shortest_timescale}).
In the first 10\,ks of the observation, strong variability is visible in the hardest energy band at around 5\,ks and 7\,ks from the start of the observation. We extract the light curve for energies above 10\,keV in the background region in order to check for the same pattern being present at hard X-rays, which is a hint for background flaring occurring during the observation. We find that an excess of up to five times the average background count rate is present at the same time as the peaks in the 4--10\,keV light curve within the first section of the hard X-ray light curve (see light-gray data in Fig.~\ref{fig:100sbinned_lcs_energydep}). Hence, we exclude these data from our study.
In addition to light curves, we extract time-resolved spectra by splitting the data into sections of 2\,ks exposure each.

Because data were obtained both in the EPIC-pn Timing and Burst modes, the uncertainties on the count rates and spectral parameters (flux and photon index) are different between the two modes. The reason for this is the different readout methods, and therefore live time, of the detector in these science modes. While it is 99.5\% for the Timing mode, in the Burst mode the detector collects data only during 3\% of the time. Accordingly, spectral parameters and count rates are less constrained when data were taken for the Burst mode.

\section{Spectral variability}\label{sec:spectral_variability}

\begin{figure*}
    \centering
    \includegraphics[width=0.99\textwidth]{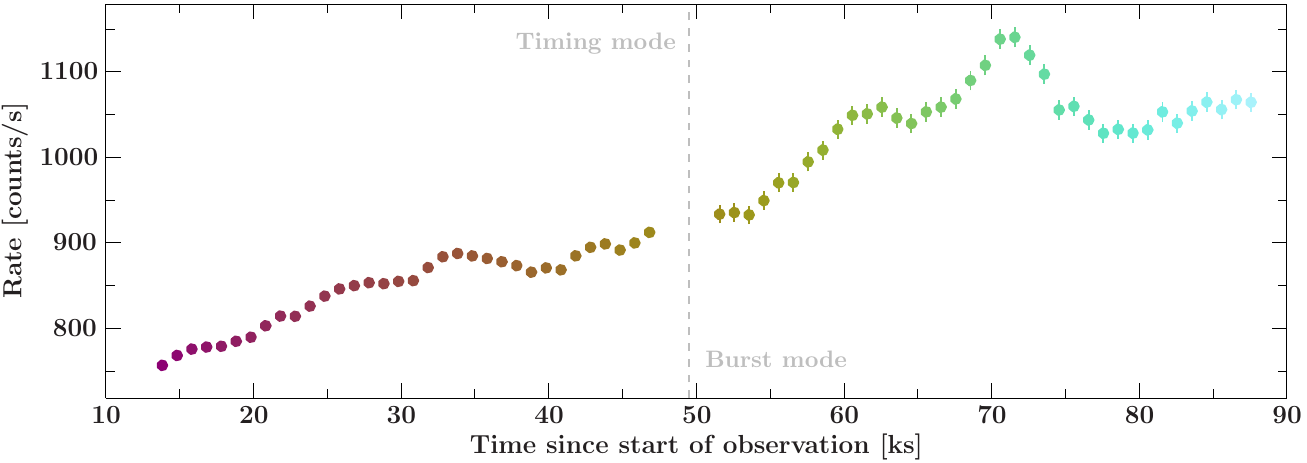}
    \caption{\xmm light curve of Mrk\,421 in 1\,ks binning, displaying the count rate for the full energy range (0.3--10\,keV). The rainbow color-coding provides an indicator of progressing time and has been applied to connect the hardness ratios in Fig.~\ref{fig:hardnessratios} to their respective occurrence in the time frame of the observation.}
    \label{fig:1ks_lightcurve}
\end{figure*}
\begin{figure*}
    \centering
    \includegraphics[width=0.99\textwidth]{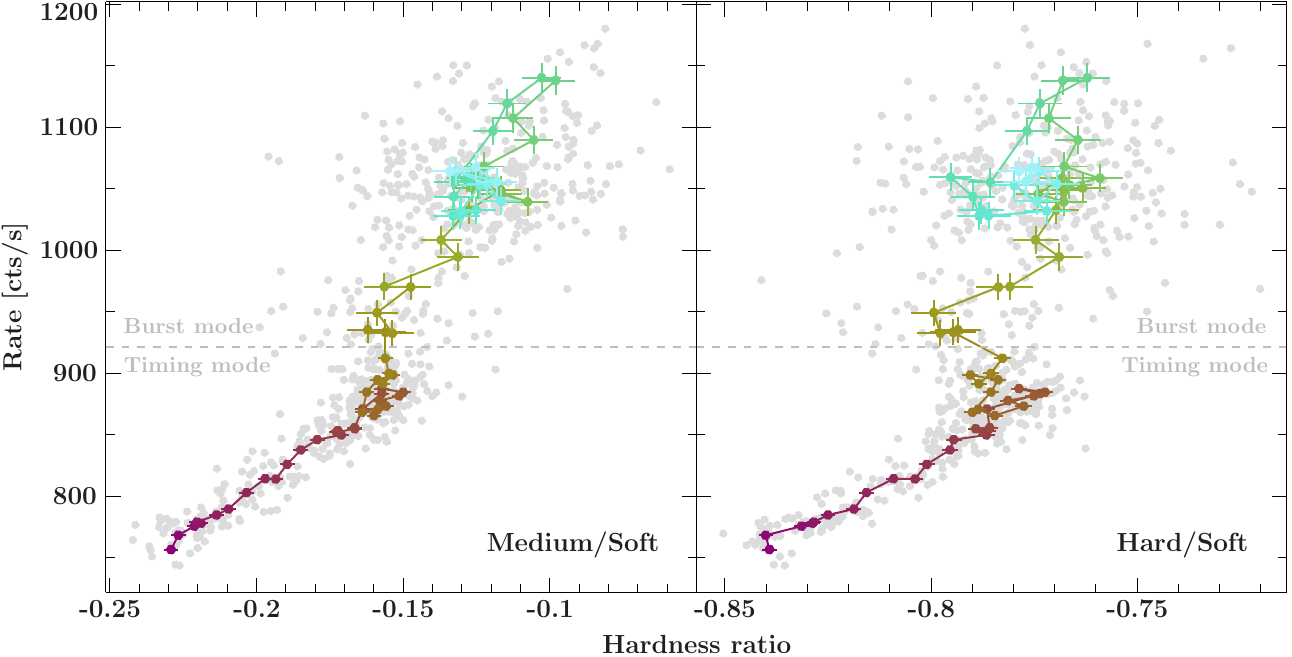}
    \caption{HIDs for medium (1--4\,keV) vs.\ soft (0.3--1\,keV) band (\textit{left}) and the hard (4--10\,keV) vs.\ soft band (\textit{right}). The grey data points are the hardness ratios of the 100-s binned light curve and display the scatter of hardness ratios on very short time scales, while the colored data points are obtained from the 1\,ks-binned time series. The colors match the time bins of the light curve displayed in Fig.~\ref{fig:1ks_lightcurve}. The sections that were taken in Timing and Burst mode, respectively, are separated by the dashed gray line.}
    \label{fig:hardnessratios}
\end{figure*}
Using this \xmm data set, it is possible to probe quasi-continuous X-ray emission on time scales longer than 20\,hrs and resolve the processes in the jet.
Ignoring the first 10\,ks due to potential background flaring, our observation covers about 74\,ks including one gap of $\sim3\,\mathrm{ks}$ length, which restricts our dataset from being strictly continuous. Nonetheless, the overall duration of the observation enables us to reveal significant spectral changes on time scales of a few hours.
As a first step, we want to use a model-independent approach to broadly investigate how the distribution of photons in different energy bands changes without any assumption of the underlying process with which the data can be modelled.
A hardness-intensity diagram (HID) is an ideal tool to quantify spectral changes during the observation, and it is defined as \citep[e.g.,][]{park2006}

\begin{equation}
\mathrm{Hardness\ ratio} = \frac{H-S}{H+S},
\end{equation}
\noindent
where $H$ and $S$ are the number of counts in the high-energy (harder) and low-energy (softer) band, respectively. By definition, the hardness value can range from $-1$, meaning that the low-energy band has a overwhelming amount of photons, to $+1$, for which significantly more photons are found in the high-energy band. Naturally, more photons are detected at lower energies, and the effective area of the EPIC favors the detection of soft photons, too, but changing emission processes or absorption can affect this order. While this quantity does not rely on any physical assumptions, it is biased by the chosen energy ranges that are compared to one another.
The uncertainty on the hardness is computed via Gaussian error propagation.

In Figure~\ref{fig:1ks_lightcurve}, we show the light curve for the 0.3--10\,keV band in 1\,ks binning, which is the same binning we choose also to create the HIDs. Figure~\ref{fig:hardnessratios} displays the HIDs for the medium (1--4\,keV) versus soft (0.3--1\,keV) band on the left, and the hard (4--10\,keV) versus soft band on the right.
As expected, the values in both HIDs are negative due to the larger amount of photon counts in the soft band compared to the medium and hard band. With a typical photon index $>2$ during an average flux state \citep[e.g.,][]{abeysekara2017,markowitz2022}, the X-ray emission from Mrk\,421 is rather soft, similar to that of most HSP sources. 

In the first part of the light curve, the flux constantly rises and the harder-when-brighter trend is visible in both HIDs. During the transition from the first to the second section of the light curve (Timing to Burst mode), the count rate seems to increase without a change of the ratio between soft and medium band (left HID), while the hard vs.\ soft HID indicates a slight decrease of photons in the 4--10\,keV band compared to those in the 0.3--1\,keV band. 
The harder-when-brighter trend continues for the second part of the light curve, and above $1000\,\mathrm{cts}\,\mathrm{s}^{-1}$, the behaviour in the medium vs.\ soft band appears to be following the same trend for both increasing and decreasing count rates. 
However, the trend seen in the HID for the hard vs.\ soft band seems to be slightly different: above $1000\,\mathrm{cts}\,\mathrm{s}^{-1}$, the hardness ratio stays at the same level during the increase of the count rate. Then, the hardness decreases when the count rate falls back to the level of ${\sim}1050\,\mathrm{counts}\,\mathrm{s}^{-1}$ after the peak flux. This behaviour creates a counter-clockwise rotation in the HID, which indicates a change to an improved cooling efficiency since the ratio of hard to soft photons stays the same during the flux increase, and becomes softer than before while the flux drops after reaching its maximum.

As a next step, we want to apply a model to the extracted spectra and determine physical parameters, that is, the slope of the underlying power law and the flux of the source. Given the extreme brightness of Mrk\,421 during the observation, spectra with a relatively short exposure of 2\,ks suffice to constrain both parameters well.
We rebin the 2\,ks-exposure spectra such that each bin has a minimum signal-to-noise ratio of 5, excluding bins below 0.3\,keV and above 10\,keV.
Each spectrum is individually fitted with an absorbed pegged power law (\texttt{tbabs*pegpwrlw}). 
For the absorption in the interstellar medium, we adopt the \texttt{vern} cross sections \citep{verner1996} and \texttt{wilm} abundances \citep{wilms2000}.
The absorption parameter is set to the Galactic hydrogren column value, $N_{\mathrm{H,Gal}}=1.33\times10^{20}\,\mathrm{cm}^{-2}$ \citep{HI4PI}, and kept frozen to avoid correlation with the photon index. The best-fit results are listed in Table~\ref{tab:mrk421_xmm_2ksfits}. 
The value of reduced $\chi^2$ for the power law model is very close to unity, indicating that the model is sufficient to describe the data sets. However, sometimes a log-parabola is found to provide a better model of the X-ray spectrum of Mrk\,421 \citep[e.g.,][]{kapanadze2016}. Here, we choose a simple power law on purpose: the curvature of the parabola ($\beta$) is correlated with the photon index ($\alpha$), and in order to describe the spectrum accurately, usually both parameters are kept free in the modelling process. In this work, we want to track the change of the spectral slope through the photon index, which is why a power law model is our favoured option to avoid cross-correlation of parameters.

We choose to visualise the time-resolved spectral changes with a so-called hysteresis curve \citep{kirk1998}.
In order to create this hysteresis curve, we use the best-fit results of the extracted spectra, which are able to verify sub-hour changes while at the same time providing adequately constrained flux and photon index values (see Table~\ref{tab:mrk421_xmm_2ksfits}).
During this 2019 June observation, Mrk\,421 presents a hard spectrum compared to spectra taken earlier during low flux states \citep[$\sim1$--$5\times10^{-10}\,\ergflux$ with power law photon indices between 2.4 and 2.8, e.g.,][]{balokovic2016}. This comes as no surprise as the `harder-when-brighter' behaviour is a well established correlation for HSP sources, and in particular Mrk\,421 \citep[e.g.,][]{zhang2019}.
A distribution of spectral indices obtained via \swift observations from 2015 until 2018 peaks at roughly 2; however, this distribution has wide tails extending to 1.6 and 2.9 \citep{kapanadze2020} demonstrating the strong and frequent changes in the steepness of the X-ray spectrum exhibited by this source.
Compared to the spectral parameters obtained by this study for similar flux levels as reported here, we find a slightly harder spectrum. However, since the data analysed by \citet{kapanadze2020} was taken with \swift and our data set was obtained through \xmm, we cannot exclude the possibility that potential systematic
differences between the two instruments may contribute to the discrepancy.

\begin{table}
\caption{Best fit results for the absorbed power law fit to the 2\,ks spectra of this \xmm observation of Mrk\,421. The flux is given for the energy range from 0.3 to 10\,keV. The results are used to display the hysteresis curve in Fig.~\ref{fig:lc_hysteresis_combo}.}
\label{tab:mrk421_xmm_2ksfits}
\centering
\begin{tabular}{c|ccc}
\hline\hline
Time & Flux & Index & Fit statistic \\
$[$ks$]$ & [$10^{-10}$ erg cm$^{-2}$ s$^{-1}$] &  & $\chi^2$/d.o.f. \\ \hline
13.3--15.3 & $16.579\pm0.001$  &  $2.058\pm0.005$  &  967.7/1044  \\
15.3--17.3 & $17.064\pm0.001$  &  $2.041\pm0.005$  &  993.7/1047  \\
17.3--19.3 & $17.271\pm0.001$  &  $2.028\pm0.005$  &  982.5/1060  \\
19.3--21.3 & $17.805\pm0.001$  &  $2.005\pm0.005$  &  1100.8/1093  \\
21.3--23.3 & $18.412\pm0.001$  &  $1.979\pm0.005$  &  1165.3/1115  \\
23.3--25.3 & $19.072\pm0.001$  &  $1.959\pm0.005$  &  1110.6/1116  \\
25.3--27.3 & $19.689\pm0.001$  &  $1.937\pm0.005$  &  1149.7/1132  \\
27.3--29.3 & $19.857\pm0.001$  &  $1.930\pm0.005$  &  1175.1/1140  \\
29.3--31.3 & $20.051\pm0.001$  &  $1.922^{+0.005}_{-0.004}$  &  1182.7/1146  \\
31.3--33.3 & $20.679\pm0.001$  &  $1.908\pm0.005$  &  1124.5/1152  \\
33.3--35.3 & $21.130\pm0.001$  &  $1.893\pm0.005$  &  1280.2/1182  \\
35.3--37.3 & $20.875\pm0.001$  &  $1.901^{+0.004}_{-0.005}$  &  1214.2/1166  \\
37.3--39.3 & $20.556\pm0.001$  &  $1.908\pm0.005$  &  1066.2/1162  \\
39.3--41.3 & $20.372\pm0.001$  &  $1.923\pm0.004$  &  1217.8/1152  \\
41.3--43.3 & $20.943\pm0.001$  &  $1.913\pm0.004$  &  1176.4/1149  \\
43.3--45.3 & $21.111\pm0.001$  &  $1.912\pm0.004$  &  1083.6/1135  \\
51.1--53.1 & $22.672\pm0.003$  &  $1.904\pm0.014$  &  444.6/505  \\
53.1--55.1 & $22.581\pm0.003$  &  $1.916^{+0.015}_{-0.014}$  &  472.1/486  \\
55.1--57.1 & $23.759\pm0.003$  &  $1.882\pm0.014$  &  483.7/522  \\
57.1--59.1 & $24.915\pm0.003$  &  $1.861\pm0.014$  &  459.5/528  \\
59.1--61.1 & $26.071\pm0.003$  &  $1.850\pm0.013$  &  478.7/532  \\
61.1--63.1 & $26.465\pm0.003$  &  $1.848\pm0.013$  &  479.1/539  \\
63.1--65.1 & $26.281\pm0.003$  &  $1.837\pm0.013$  &  465.9/541  \\
65.1--67.1 & $26.511\pm0.003$  &  $1.843\pm0.013$  &  509.4/540  \\
67.1--69.1 & $27.363\pm0.004$  &  $1.827\pm0.012$  &  569.9/553  \\
69.1--71.1 & $28.650\pm0.004$  &  $1.818\pm0.012$  &  582.3/559  \\
71.1--73.1 & $28.737\pm0.004$  &  $1.826\pm0.012$  &  476.9/561  \\
73.1--75.1 & $26.757\pm0.003$  &  $1.862\pm0.013$  &  470.4/540  \\
75.1--77.1 & $25.985\pm0.003$  &  $1.876\pm0.013$  &  514.1/529  \\
77.1--79.1 & $25.537\pm0.003$  &  $1.868\pm0.013$  &  492.7/531  \\
79.1--81.1 & $25.602\pm0.003$  &  $1.862\pm0.013$  &  512.5/538  \\
81.1--83.1 & $26.219\pm0.003$  &  $1.850\pm0.012$  &  516.1/539  \\
83.1--85.1 & $26.510\pm0.003$  &  $1.852\pm0.013$  &  490.3/542  \\
85.1--87.1 & $26.577\pm0.003$  &  $1.851^{+0.013}_{-0.012}$  &  492.9/545  \\
\hline
\end{tabular}
\end{table}

\begin{figure*}
    \begin{minipage}{0.49\textwidth}
    \includegraphics[width=\textwidth]{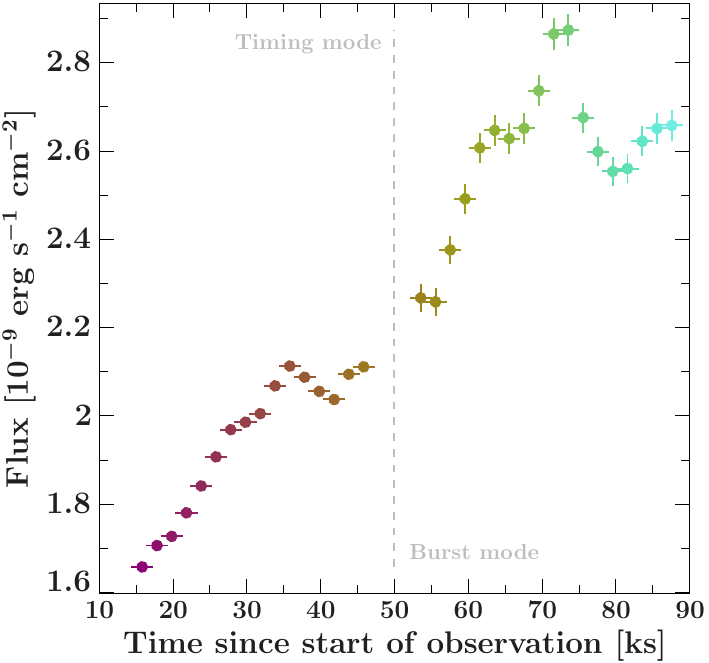}
    \end{minipage}
    \hfill
    \begin{minipage}{0.48\textwidth}
    \includegraphics[width=\textwidth]{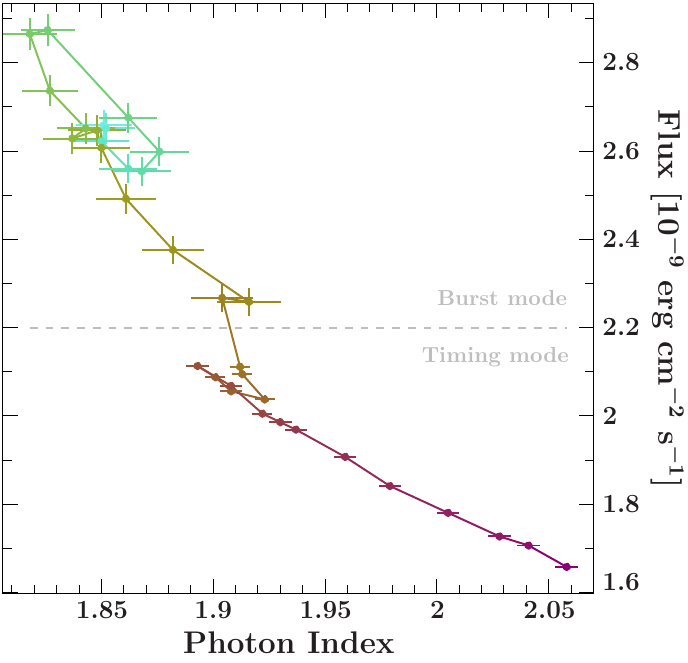}
    \end{minipage}
    \caption{\xmm light curve (\textit{left}) in 2\,ks binning and corresponding hysteresis curve (\textit{right}). The flux is given for the 0.3--10\,keV band. The color-coding is used to enable an easier connection of the light curve to the data points of the hysteresis curve. The sections that were taken in Timing and Burst mode, respectively, are separated by the dashed gray line, and the smaller uncertainties for flux and photon indices measured in the Timing mode are due to the significantly higher live-time and therefore photon collecting ability of this observing mode. During both the Timing and the Burst mode, the increase and decrease of a (localized) peak in the light curve can be observed at 35\,ks and 70\,ks, respectively. While for the peak in the Timing mode section of the light curve the path on the hysteresis curve is identical for rise and decay, a hysteresis loop pattern is visible for the peak in the Burst mode section.}
    \label{fig:lc_hysteresis_combo}
\end{figure*}

The resulting hysteresis curve and an accompanying light curve in 2\,ks-binning are shown in Fig.~\ref{fig:lc_hysteresis_combo}.
During the majority of the observing time, the X-ray flux of Mrk\,421 increases. In addition to the increasing flux, the spectrum becomes significantly harder, hence the movement from the lower right corner to the upper left corner in the hysteresis plot. This harder-when-brighter behaviour is also seen in the model-independent HID. The overall spectral hardening takes place within less than 1\,day, with the continuous increase being roughly 65\,ks long.
For the first section of the light curve, the flux steadily increases most of the time, while the photon index initially decreases, before behaving in the opposite way: increasing while the flux gets slightly lower (e.g., the small, intermediate peak seen at $\sim$35\,ks). Here, no loop can be observed.
In the second section of the light curve, the flux peaks at $\sim$70\,ks, before it visibly drops. For this behaviour, a clear clockwise rotation can be observed in the hysteresis curve, which indicates that the cooling time scale dominates the underlying physical processes, as the high-energy particles cool efficiently and the accelerated particle population interacts with the low-energy photons later, i.e., showing a `soft lag' \citep[c.f.][]{kirk1998}. The `soft lag'-behaviour is a common occurrence in the X-ray emission of HSP-type blazars \citep[e.g.,][]{falcone2004,wang2018}, and has also been seen in the optical \citep{agarwal2021}.
For Mrk\,421, both clockwise and anti-clockwise rotating hysteresis curves have been found in the past: data taken with \xmm (0.5--10\,keV) as well as \textit{RXTE}/PCA (2--60\,keV) revealed opposing behaviour, in some cases for observations spaced just a few days apart \citep{ravasio2004, cui2004, abeysekara2017}. The dissimilar evolution of flux and spectral shapes indicates that we might see different types of flares, or different stages within flaring events. 
However, observing times often only cover a relatively short period of time (usually hours) compared to the length of an activity phase, and since strictly continuous monitoring over several days with an instrument like \xmm is currently not available, it is only possible to work with data that covers only a small segment, which is biased towards flaring states due to target-of-opportunity based observation strategies.

\section{Time series analysis}\label{sec:time_series_analysis}
While we are able to track down spectral variations down to a few kilo-seconds for this data set, the time resolution of the EPIC-pn onboard \xmm enables us to study the variability properties of sources down to minutes or even seconds. We utilize the light curves to derive energy-resolved values for the fractional variability amplitude (Sect.~\ref{sec:frac_var}), to search the shortest source-intrinsic variability signatures (Sect.~\ref{sec:search_shortest_timescale}), and to determine whether or not a lag is present between the different X-ray energy bands (Sect.~\ref{sec:lag_search}). 
\subsection{Fractional variability amplitude}\label{sec:frac_var}
The fractional variability amplitude ($F_{\mathrm{var}}$) is a quantity commonly used to derive a measure of the variability in a time series normalized to a source's average flux. $F_{\mathrm{var}}$ is a useful tool for unevenly sampled data and has therefore often been used for the analysis of data obtained in multi-wavelength campaigns. Here, we use it to provide values that can be compared to other work that has been done on similar data sets.
It was defined by \citet{vaughan2003} as 

\begin{equation}F_{\mathrm{var}}=\sqrt{\frac{\sigma_{\mathrm{nxs}}}{\overline{x}^2}}=\sqrt{\frac{S^2-\overline{\sigma^2_{\mathrm{err}}}}{\overline{x}^2}},
\end{equation}
\noindent
where the excess variance $\sigma_{\mathrm{nxs}}$ \citep{nandra1997,edelson2002} is divided by the average flux of the time span taken into account. $S^2$ describes the general variance, and $\overline{\sigma^2_{\mathrm{err}}}$ is the expected measurement uncertainty.
The uncertainty of $F_{\mathrm{var}}$ can be calculated via

\begin{equation}
    \mathrm{err}(F_{\mathrm{var}}) = \sqrt{\left(\sqrt{\frac{1}{2N}}\frac{\overline{\sigma^2_{\mathrm{err}}}}{\overline{x}^2 F_{\mathrm{var}}}\right)^2 + \left(\sqrt{\frac{\overline{\sigma^2_{\mathrm{err}}}}{N}}\frac{1}{\overline{x}}\right)^2}\,,
\end{equation}
\noindent
where $N$ is the number of all measurements.
We compute $F_{\mathrm{var}}$ for the 100s-binned light curve and list the results in Table~\ref{tab:mrk421_fracvar_tvar}.
For Mrk\,421, the fractional variability amplitude at X-ray energies ranges typically from 40\% to 80\% when covering time ranges of several days to weeks or months \citep[e.g.,][]{giebels2007,horan2009,balokovic2016,ahnen2016,acciari2020}, however, our data cover a time range of less than a day, resulting in a lower fractional variability amplitude. A systematic study by \citet{schleicher2019}, which is based on FACT monitoring data of blazars, has shown that the size of the analysed data set and the time span which a data set covers can strongly influence the fractional variability amplitude obtained.
Therefore, we compare our results to a study of 25 observations (average duration of $\sim38$\,ks) of Mrk\,421 with \xmm, which covered all observations between 2000 and 2017 with a minimum duration of 10\,ks \citep{noel2022}. 
The data set we use to determine $F_{\mathrm{var}}$ contains a total of 726 data points.
The value of $F_{\mathrm{var}}$ for this observation is significantly above the average value of $F_{\mathrm{var}}$ derived from the systematic study.
While \citet{noel2022} have potentially found a slight positive correlation between the variability of a source and its flux exists, the length of an observation also influences fractional variability towards higher values. This connection indicates the presence and dominance of red noise in the light curves of Mrk\,421.

\begin{table}
	\caption{Fractional variability amplitude, $F_{\mathrm{var}}$, and minimum variability time scale, $\tau_{\mathrm{var}}$, computed for the full energy range as well as the three energy sub-bands.}
	\begin{center}
	\begin{tabular}{lcc}
	\hline
	\hline
	Energy band	&  $F_{\mathrm{var}}$ [\%] &  $\tau_{\mathrm{var}}$ [s] \\
	\hline	
	0.3-10\,keV 	& $11.3\pm0.1$ &  $955\pm257$ \\

	\hline
	0.3-1\,keV	 	& $8.6\pm0.1$ & $585\pm141$ \\ 
	1-4\,keV		& $14.5\pm0.1$ & $734\pm263$ \\
	4-10\,keV		& $15.9\pm0.3$ & $297\pm115$ \\
	\hline  
	\end{tabular}
	\end{center}
	\label{tab:mrk421_fracvar_tvar}
\end{table}
In addition, the values for $F_{\mathrm{var}}$ exhibit a clearly visible energy dependence, with the fractional variability amplitude being almost double for the hard vs.\ soft band.
Prior studies have found the same trend that the value of $F_{\mathrm{var}}$ increases with X-ray energy for Mrk\,421 \citep{giebels2007, markowitz2022, schleicher2019, arbet-engels2021, noel2022}, and even for radio-loud AGN in general \citep{peretz2018}.

\subsection{Search for shortest time scales}\label{sec:search_shortest_timescale}
By searching for the shortest time scales on which a source exhibits variability, we can obtain knowledge of the underlying processes and potential structure within the emission region.
Our first approach to search for short-time variability is computing the flux variability time scale, also known as e-folding time.
\citet{burbidge1974} defined an estimate of the flux-normalized (or weighted) variability time scale via

\begin{equation}
\tau_{\mathrm{var}} = \left\vert \frac{\Delta t}{\Delta \mathrm{ln}F} \right\vert = \left\vert \frac{t_1-t_2}{\mathrm{ln}(F_1/F_2)} \right\vert,
\end{equation}
\noindent
with $\Delta t$ being the time interval between two selected flux measurements, $F_1$ and $F_2$.
An uncertainty for $\Delta\tau_{\mathrm{var}}$ can be derived via standard error propagation \citep[see also][]{bhatta2018}, which yields

\begin{equation}
\Delta\tau_{\mathrm{var}} \simeq \sqrt{\frac{F^2_1\Delta F^2_2 + F^2_2\Delta F^2_1}{F^2_1 F^2_2 (\mathrm{ln}[F_1/F_2])^4}}\,\Delta t.
\end{equation}

We compute $\tau_{\mathrm{var}}$ for all combinations of two data points in the 100s-binned light curve. From those results, we select the smallest value that fulfils the condition of $\vert F_1-F_2\vert > \Delta F_1+\Delta F_2$, which ensures that the variation lies outside of the flux uncertainties.
The resulting values are listed in Table~\ref{tab:mrk421_fracvar_tvar}. 

We find a value of $\tau_{\mathrm{var}}$ of $\sim1\,\mathrm{ks}$ for the full energy range, which translates to variability on a time scale of roughly 16 minutes. 
The value of $\tau_{\mathrm{var}}$ for the sub-bands is even smaller, with a fastest variability time scale of ${\sim}5\,\mathrm{min}$ at the highest energies.

In previous studies of Mrk\,421 that have used $\tau_{\mathrm{var}}$ as a tool, similar time scales for the full energy band have been found \citep{yan2018,chatterjee2021,noel2022}.
\citet{noel2022} do not find any clear correlation of $\tau_{\mathrm{var}}$ with the flux state of Mrk\,421. Another study which did not include data from Mrk\,421, however, reported a tentative correlation of $\tau_{\mathrm{var}}$ with the mean flux, for which the fastest variability occurs during the lowest flux state of a source \citep{bhatta2018}. This study analysed \textit{NuSTAR} data of 13 blazars, both of the FSRQ and the BL Lac type. Since the observation of Mrk\,421 presented in this work reveals rapid variability during a very high X-ray flux state, our findings do not match their proposed correlation.

\begin{figure}
    \centering
    \includegraphics[width=\columnwidth]{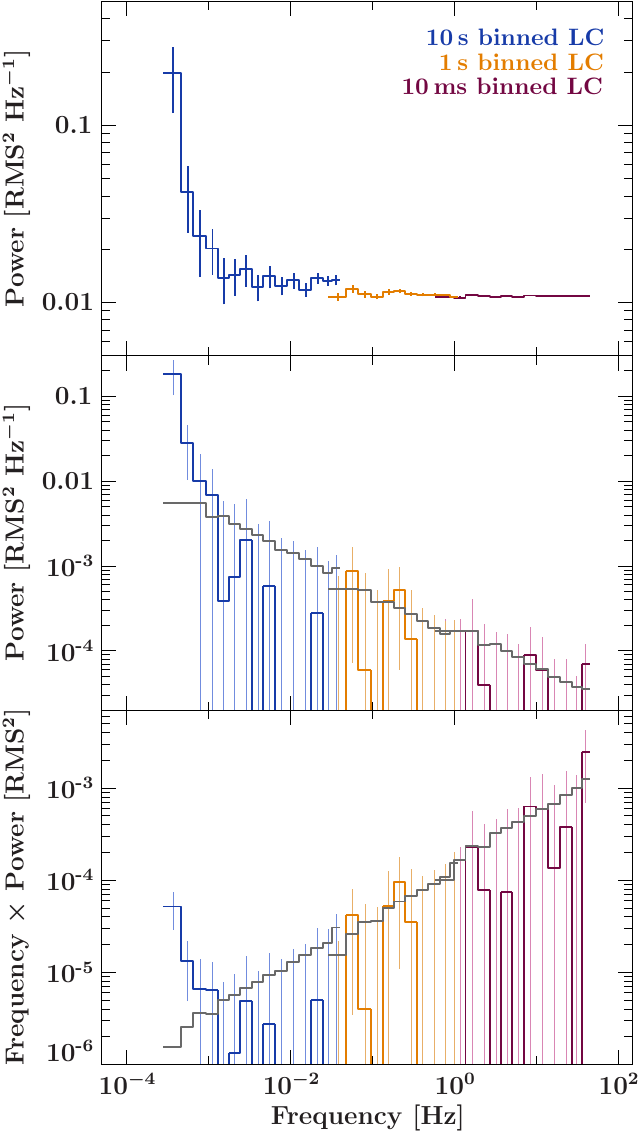}
    \caption{Combined power spectral densities for Timing mode light curves with 10\,s, 1\,s, and 10\,ms binning (\textit{top panel}), the Poisson noise subtracted version of these PSDs (\textit{middle panel}), and the Poisson-reduced PSDs multiplied by frequency (\textit{bottom panel}). The gray line shows the upper $1\sigma$ envelope for the uncertainty of the Poisson reduction.}
    \label{fig:psds}
\end{figure}

A more refined method to study variability in any given time series is using a discrete Fourier transformation $X_j$ in order to compute a power spectral density (PSD), which is defined as

\begin{equation}
    P_j=X_j\cdot X^*_j,
\end{equation}
\noindent
with $X^*_j$ being the complex conjugate. Since the Timing mode is able to provide a large live-time in combination with a high time resolution of less than 1\,ms, we will use the data taken in this mode, and use light curves with different binning to create a combined PSD that spans a large frequency range from $10^{-4}$ to $>10\,\mathrm{Hz}$.
We use a light curve with a binning of 10\,ms, and in order to stay consistent, we create light curves with a binning of 1\,s and 10\,s from this initial light curve.
We ignore all values for which either the rate is \texttt{nan}, or the fractional exposure is lower than 90\%. 
The count rate is steadily rising in the light curve, and therefore does not initially fulfill the condition for stationarity, which is necessary for a timing analysis via a PSD. In order to account for this behaviour, we fit the 10\,s light curve with a linear function, $at+b$, and subtract a straight line with parameters from the best fit results from each light curve. Then, we add the average count rate determined by a linear fit of each light curve, respectively, in order to get the appropriate normalization.

We compute periodograms; to reduce the scatter inherent in single periodograms, we perform averaging across multiple periodograms derived from multiple light curve segments. We optimise the segment length to cover the maximum possible time range that doesn't include any gaps, while also producing a reasonable amount of segments.
For the 10\,ms, 1\,s, and 10\,s binned light curves we produce 4084, 419, and 6 segments that we create periodograms for and average over, respectively.
In addition, we apply a frequency rebinning of $\Delta f/f=0.4$. 
We perform a Monte Carlo simulation for each PSD to estimate the level of the Poisson noise contribution to the PSD based on the measured light curves by taking into account the statistical properties, the linear increase, and occurrences of gaps, and compute PSDs in the same way as has been done for the real data.
We use the results from the simulation to subtract the Poisson noise from the PSDs in order to determine up to which frequency, and therefore time scales, variability is present in the light curve of Mrk\,421.
In Figure~\ref{fig:psds}, we show the computed PSDs in the top panel, and the Poisson noise-subtracted PSDs in the other two panels, with the bottom one showing the PSDs multiplied by frequency.
The upper $1\sigma$ uncertainty envelope is indicated by the gray line, and makes visible where power remains above the Poisson noise level. As can be seen clearly in the middle and lower panel in Fig.~\ref{fig:psds}, no variability remains at frequencies above $\sim10^{-3}\,\mathrm{Hz}$. This translates to a time scale of about 1\,ks, which is in agreement with our calculation of $\tau_{\mathrm{var}}$, as well as other studies \citep{abeysekara2017,yan2018,chatterjee2021,noel2022}, and seems to be a characteristic time scale for Mrk\,421.

\subsection{Search for energy-related time lags}\label{sec:lag_search}

\begin{figure}
    \includegraphics[width=\columnwidth]{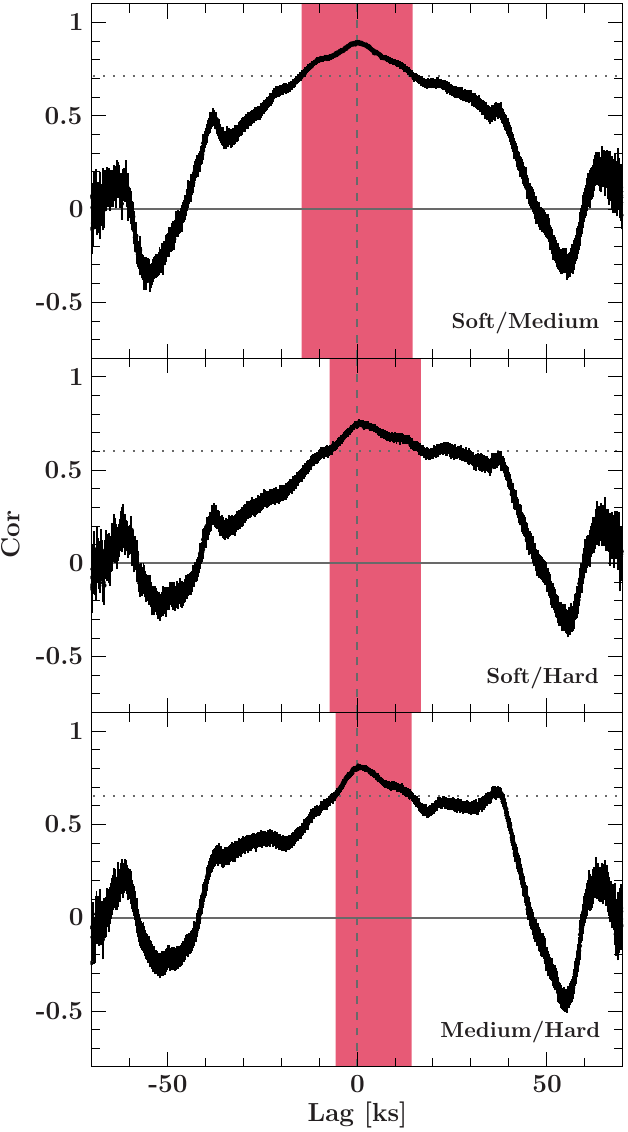}
    \caption{zDCFs computed for the energy-resolved \xmm light curves of Mrk\,421. From top to bottom, the panels display the correlation between the soft and medium, soft and hard, and medium and hard energy band. The dashed line marks the zero lag, and the threshold for $0.8\times r_{\mathrm{max}}$ is marked by the gray dotted horizontal line. Values that fulfil $>0.8\times r_{\mathrm{max}}$ and are taken into account for computing $\tau_{\mathrm{cent}}$ are highlighted by the red-shaded area in the background.}
    \label{fig:dcfs}
\end{figure}
The detection of a presence, or a lack, of time lags between two energy bands can reveal information about acceleration vs. cooling time scales during flares. Studies searching for lags between the soft and hard X-ray band have yielded different results as Mrk\,421 seems to present an inconsistent behaviour with showing no lag during some times, but also exhibiting either a positive or negative lag that can be as large as a few kiloseconds at other times \citep[e.g.,][]{takahashi1996,fossati2000,takahashi2000,tanihata2001,ravasio2004,lichti2008,arbet-engels2021,markowitz2022}.
A tool that is commonly used for lag determination is the discrete correlation function \citep[DCF;][]{edelson1988}. It is ideal for irregularly sampled data sets and data that show gaps, such as our \xmm data set. In this work, we are using the z-transformed DCF by \citet{alexander1997,alexander2013}, which alleviates biases associated with the classic DCF.
We use the 100s-binned light curves to resolve potential short-time lags, and compute the zDCFs for the combinations soft vs.\ medium band, soft vs.\ hard band, and medium vs.\ hard band. Our maximum lag is the length of the light curve (75\,ks) with a binning of 100\,s for the zDCF curve and we ensure that the binning is centered around a lag value of zero seconds.
The resulting three zDCFs are shown in Fig.~\ref{fig:dcfs}. The correlation for the two combinations, 0.3--1\,keV vs. 4--10\,keV (Soft/Hard) and 1--4\,keV vs. 4--10\,keV (Medium/Hard), appears more asymmetric towards positive lag times than the correlation of the 0.3--1\,keV vs. 1--4\,keV (Soft/Medium) bands.
In order to determine the lag, we use two techniques: (1) determining the lag $\tau_{\mathrm{peak}}$ at the position of the maximum correlation factor $r_{\mathrm{max}}$, and (2) averaging over all points with correlation coefficients $>0.8\,r_{\mathrm{max}}$ in order to obtain a central time lag value $\tau_{\mathrm{cent}}$.
\begin{table}
	\caption{Results from applying the discrete correlation function to the 100s-binned light curves. Lags are defined such that positive values indicate harder photons leading softer photons.}
	\vspace{-0.3cm}
	\begin{center}
	\begin{tabular}{l|ccc}
	\hline
	\hline
	DCF	&  $r_{\mathrm{max}}$ &  $\tau_{\mathrm{peak}}$ [ks]  &  $\tau_{\mathrm{cent}}$ [ks]\\
	\hline	
	Soft/Medium	 	& 0.894$^{+0.007}_{-0.008}$ & $0.3\pm0.7$ & $0.0\pm8.5$\\ 
	Soft/Hard		& $0.756\pm0.016$ & $0.5^{+1.6}_{-0.9}$ & $4.8\pm7.0$\\
	Medium/Hard		& 0.814$^{+0.012}_{-0.013}$ & $0.3\pm^{+1.4}_{-0.6}$ & $4.4\pm5.9$\\
    \hline
	\end{tabular}
	\end{center}
	\label{tab:mrk421_dcfvalues}
\end{table}
The results for both approaches are listed in Table~\ref{tab:mrk421_dcfvalues}. The error on $\tau_{\mathrm{peak}}$ is calculated via the zDCF using a maximum likelihood method.
We note that for the Soft/Hard and Medium/Hard correlation, the curves reach briefly above the threshold of $>0.8\,r_{\mathrm{max}}$ at lags of $\sim23$\,ks and $\sim37$\,ks, respectively. We exclude these parts in order to have a continuous portion of the zDCF to fit.
We consistently find that peak lag values are slightly positive (suggesting harder photons leading softer ones), although values are formally consistent with zero lag. Across all light curves, we can safely exclude that soft photons lead hard photons by $\geq$\,0.3--0.4\,ks. We also exclude that hard photons lead soft ones by values $\geq$\,1.0, 2.1, and 1.7 ks in the Medium/Soft, Hard/Soft, and Hard/Medium correlations, respectively.
The tendency towards a soft lag in relation to the hard X-ray band is also visible in the hysteresis curve in Fig.~\ref{fig:lc_hysteresis_combo}, where a clock-wise loop appears that describes the last 30\,ks of the observation.

In general, Mrk\,421 seems to exhibit both soft, hard, and also no lags in the X-ray band.
While the X-ray emission in different energy bands is typically correlated, this correlation is often found to be consistent with a time lag of zero, as was reported based on data obtained by \textit{NuSTAR} \citep[3--10\,keV vs. 10--79\,keV;][]{pandey2017}, \xmm \citep[0.3--2\,keV vs. 2--10\,keV][]{noel2022}, \textit{Chandra} \citep[0.3--2\,keV vs. 2--10\,keV;][]{aggrawal2018}, and ASTROSAT \citep[0.7--7\,keV vs. 7--20\,keV;][]{das2023}.
In other cases it was found that the lower-energy emission leads the higher-energy emission (i.e., a hard lag) based on data obtained with BeppoSAX \citep[0.1--1.5\,keV vs. 3.5--10\,keV / 0.1--2\,keV vs. 2--10\,keV;][respectively]{fossati2000,zhang2002}, ASTROSAT \citep[0.6--0.8\,keV vs. several sub-bands reaching up to 7\,keV;][]{markowitz2022}, and INTEGRAL \citep[20--40\,keV vs. 40--100\,keV;][]{lichti2008}, with the tendency to be observed during flaring events.
A soft lag was reported based on data from an \xmm observation \citep[0.6--0.8\,keV vs. 4--10\,keV][]{zhang2010}, as well as an ASCA observation \citep[0.5--1\,keV vs. 2--7.5\,keV][]{takahashi1996}.
Some studies found a time lag being present only for a part of the analysed light curve \citep{brinkmann2005, markowitz2022}, indicating that the involved emission processes can change relatively quickly.
Due to the uncertainties on our derived values, which are determined by the standard deviation of $\tau_{\mathrm{cent}}$, our results are consistent with a zero time lag, even though they tentatively hint at a soft lag.

\section{Discussion \& Conclusions}\label{sec:discussion}
We analysed a target-of-opportunity \xmm observation of Mrk\,421 during a very bright X-ray flare that occurred simultaneously to VHE \gm-ray activity in June 2019. Our spectral analysis reveals typical harder-when-brighter behaviour. From the clock-wise hysteresis loop and a tentative soft lag, which is found by employing the z-transformed discrete correlation function, we conclude that the electron cooling dominated over the acceleration process during the flare.
Our timing analysis determines a fractional variability amplitude above the average level of single pointed observations and rapid variability down to time scales of $\sim300$\,s at the highest energies, and 1\,ks in the full energy range.
The fastest variability timescales that we find can give as an estimation of the size of the emission region for such rapid variability.
In our calculations, we use $z=0.031$, and $M_{\mathrm{BH}}=1.9\times10^8$ M$_{\odot}$ \citep{woo2002}, which yields a Schwarzschild radius of 1\,$R_{\mathrm{S}}=2.9\times10^{13}\,\mathrm{cm}$.
If we naively assume that the emission region spreads over the entire width of the jet, and that the jet employs a conical shape \citep[which has been found to be true for most HSP and radio-loud quasars up to several kpc, e.g.,][]{pushkarev2017}, we can compute the distance of the emission region from the jet base. For an opening angle $\alpha$, this distance of the jet base is given by

\begin{equation}
d_{\mathrm{jb}}=\frac{d_{\mathrm{r}}}{2\,\mathrm{tan}\,(\alpha/2)}\,,
\end{equation}
\noindent
with $d_{\mathrm{r}}$ being the upper limit for the size of the emission region (based on the light crossing time of the fastest variability time scale), and $\alpha$ being the opening angle.
The intrinsic opening angles of blazar jets have been found to lie between $0.1^{\circ}$ and $9.4^{\circ}$, with the median being $1.2^{\circ}$ for sources detected by the Large Area Telescope onboard the \fermi satellite \citep{pushkarev2017}, which we choose for Mrk\,421. 

The Doppler factors of blazars can be estimated via several different methods \citep[see, e.g.,][for a review]{liodakis2015}, but often these yield only lower limits, or even contradictory results for the same sources.
Studies modelling broadband SEDs, in particular those that contain a data set representing a flaring state, usually require large values of $\delta$ in order to explain the high luminosity at $\gamma$-ray energies.
For example, to model the high-activity states of Mrk\,421, Doppler factors in a range from 20 to 48 have been derived \citep{tavecchio1998,donnarumma2009,kapanadze2016,banerjee2019,zheng2021,markowitz2022}.
However, other work relying on radio data and measuring the variability Doppler factor of Mrk\,421 found $\delta=1.7$ \citep{liodakis2017b}, and $\delta=2.03$ \citep{liodakis2018}, and a lower limit of $\delta>2.47$ based on $\gamma$-ray opacity was determined by \citet{pei2020}.
In Table~\ref{tab:results_er_quantities}, we therefore list the computed values for 
the time scales in the jet frame ($t_{\mathrm{jet}}$), the size of the emission region ($d_{\mathrm{r}}$), and the distance of the emission region to the jet base ($d_{\mathrm{jb}}$) only in relation to $\delta$.
We derive these parameters for the fastest variability time scales in the full and the highest energy band, as well as based on the long-term variability defined by the duration of the X-ray flux increase.

\begin{table*}
\caption{Transferred time scales in the jet frame ($t_{\mathrm{jet}}$), size of the emission region ($d_{\mathrm{r}}$), and distance of the emission region to the jet base ($d_{\mathrm{jb}}$) under the assumption that the region covers the full width of the jet. We assume $M_{\mathrm{BH}}=1.9\times10^8$ M$_{\odot}$, $\alpha=1.2^{\circ}$, and $z=0.031$. Using those parameters, 1\,R$_{\mathrm{S}}=2.9\cdot10^{13}\,\mathrm{cm}$.}
\begin{center}
\begin{tabular}{lccccc}
\hline\hline
& Energy band & $t_{\mathrm{obs}}$ & $t_{\mathrm{jet}}$ & $d_{\mathrm{r}}$ & $d_{\mathrm{jb}}$  \\\hline
\multirow{2}{*}{Rapid} &  0.3--10\,keV   & $955\pm257$\,s     &  $(15\pm4)\delta$\,min    & $(0.49\pm0.13)\delta$\,R$_{\mathrm{S}}$   & $(23\pm6)\delta$\,R$_{\mathrm{S}}$ = $(4\pm1)\delta\times10^{-4}$\,pc  \\ 
 &  4--10\,keV      & $297\pm115$\,s     &  $(5\pm2)\delta$\,min   & $(0.15\pm0.06)\delta$\,R$_{\mathrm{S}}$    & $(7\pm3)\delta$\,R$_{\mathrm{S}}$ = $(1.3\pm0.5)\delta\times10^{-4}$\,pc\\ \hline
Slow &  0.3--10\,keV      & $57\pm1$\,ks     &  $(15.4\pm0.3)\delta$\,hours   &  $(29.4\pm0.5)\delta$\,R$_{\mathrm{S}}$  & $(1402\pm25)\delta$\,R$_{\mathrm{S}}$ =  $(2.57\pm0.05)\delta\times10^{-2}$\,pc  \\ \hline
\end{tabular}
\end{center}
\label{tab:results_er_quantities}
\end{table*}

In a similar way, we can estimate the magnetic field present in the emission region, both for the entire slowly-varying envelope as well as for the region where the rapid variability is created.
Given that no lag is observed in the light curve, we know that cooling alone does not dominate the observed variability. Still, we can assume that a significant amount of variability is created through synchrotron cooling, for which the cooling time scale for an electron in the co-moving frame is given as \citep[e.g.,][]{tavecchio1998}

\begin{equation}\label{eq:coolingtime}
    t_{\mathrm{cool}}=\frac{6\pi m_e c}{\sigma_{\mathrm{T}}\gamma B^2}
\end{equation}
\noindent
where $m_e$ is the electron rest mass, $\sigma_{\mathrm{T}}$ is the Thomson cross section, $\gamma$ is the Lorentz factor of the ultra-relativistic electron, and $B$ is the magnetic field in the co-moving frame of the jet.
We can rearrange Eq.~\ref{eq:coolingtime} to solve for the magnetic field $B$. By applying $t_{\mathrm{cool}}=\delta t_{\mathrm{cool,obs}}/(1+z)$, we obtain

\begin{equation}
    B = \sqrt{\frac{6\pi(1+z)m_e c}{\sigma_{\mathrm{T}}\delta\gamma t_{\mathrm{cool,obs}}}}.
\end{equation}
\noindent
Given that we have no means to measure the Lorentz factor $\gamma$, we replace the Lorentz factor through its relation with the peak synchrotron emission $E_{\mathrm{sync}}$ energy instead. Using the expression by \citet{nalewajko2011}, which is given for an averaged magnetic pitch angle, 

\begin{equation}
    E_{\mathrm{sync}}=2\cdot10^{-11}\delta B\gamma^2(1+z)^{-1}\,\mathrm{keV}
\end{equation}
\noindent
we can compute an estimate for the magnetic field $B$ via

\begin{equation}
    B=\left(\frac{6\pi m_e c}{\sigma_{\mathrm{T}}t_{\mathrm{cool,obs}}}\right)^{2/3}\left(\frac{2\cdot10^{-11}\,\mathrm{keV}(1+z)}{\delta E_{\mathrm{sync}}}\right)^{1/3} \,\mathrm{G}\,.
\end{equation}
\noindent
We use $E_{\mathrm{sync}}=5$\,keV as the average energy of the 0.3--10\,keV band.
For the large region that shows the slowly varying envelope and an increase time of roughly 57\,ks in the full energy band (0.3--10\,keV), we obtain a lower limit on the magnetic field of $B\geq0.09\delta^{-1/3}$\,G, which is similar to the findings of \citet{markowitz2022}. 
Using the shortest variability time for the 0.3--10\,keV band, we estimate 
$B\geq1.4\delta^{-1/3}$\,G, and for 4--10\,keV (using $E_{\mathrm{sync}}=7$\,keV for the average energy of that band) $B\geq2.7\delta^{-1/3}$\,G.

Considering the range of values for $\delta$ obtained through multi-wavelength studies of the source, we can test if the peak luminosity of the high-energy component, which is created through synchrotron-self Compton scattering, is at the expected value. For this, we make the simple assumption that the VHE \gm-ray flux obtained with FACT is covering the decline of the SSC component close to the peak and use equation (2.4) from \citet{ghisellini1996} to solve for a lower limit of the expected luminosity of the Inverse Compton component, $L_{\mathrm{C}}$. 
For the following calculations, we adopt a flat cosmology with $H_0= 67.8\,\mathrm{km}\,\mathrm{s}^{-1}\,\mathrm{Mpc}^{-1}$, $\Omega_{\lambda}=0.692$, and $\Omega_{\mathrm{M}}=0.308$ \citep{planck2016}.
Using a spectral index of $\alpha=3.35$ for the \gm-ray spectrum, which is the average slope of FACT spectra reported by \citet{arbet-engels2021},
and a flux of $5.1\times10^{-11}$\,ph\,cm$^{-2}$\,s$^{-1}$, we obtain $L_{\gamma}\geq2.5\times10^{44}$\,erg\,s$^{-1}$. 
To estimate the luminosity of the synchrotron peak, we use the spectral parameters during the highest X-ray flux measured with \xmm (see Table~\ref{tab:mrk421_xmm_2ksfits}), which yields L$_{\mathrm{syn}}\sim6.9\times10^{45}\,\mathrm{erg}\,\mathrm{s}^{-1}$.
For a range of Doppler factors from 20 to 48, we find $1.6\times10^{44}\,\mathrm{erg}\,\mathrm{s}^{-1}\leq\mathrm{L}_{\mathrm{C}}\leq1.7\times10^{46}\,\mathrm{erg}\,\mathrm{s}^{-1}$. Since the luminosity of the SSC component during our observation is at the lower end of that range, we adopt a Doppler factor of $\delta=40$ for the continuing discussion regarding the origin of the X-ray emission in the jet.

If we use the shortest time scales to constrain the full size of the emission region responsible for the overall flaring activity, we find that it has to be located relatively close to the central engine, since $d_{\mathrm{jb}}\leq280\,\mathrm{R}_{\mathrm{S}}$ and $d_{\mathrm{jb}}\leq920\,\mathrm{R}_{\mathrm{S}}$ using $\tau_{\mathrm{var}}$ measured for the 4--10\,keV band and 0.3--10\,keV band, respectively.
However, in addition to the rapid variability embedded in the X-ray light curve, Mrk\,421 exhibits a steady flux increase from $t=13$\,ks to $t=70$\,ks (see, e.g., Fig.~\ref{fig:100sbinned_lcs_energydep}). Using a time scale of $\sim57\pm1$\,ks for the light crossing time of the emission region in the observers frame, we conclude that the emission region responsible for the steady increase, if covering the entire width of the jet, would likely lie at a distance of $\sim$1\,pc from the central engine.
The size of this larger emission region is comparable to what has been found by \citet{tramacere2022} for Mrk\,421, based on a model involving an expanding adiabatic blob, and which is able to produce the frequently observed time lags between GeV \gm-ray and radio emission from blazars.
However, neither this model nor the standard shock acceleration model are able to explain the observed rapid variations, requiring us to turn to other, complementary models, such as a `minijets-in-jet'\citep[e.g.,][]{giannios2013,shukla2020} or a `spine-sheath' / 'current sheet' scenario, in which (kink) instabilities lead to magnetic reconnection events \citep[e.g.,][]{uzdensky2010,sironi2015,petropoulou2016,bodo2021,zhang2022}. Both of these models were developed to explain rapid \gm-ray variability on time scales of a few minutes, which was observed for a handful of sources during extremely bright flares \citep{aharonian2007,albert2007,aleksic2014,ackermann2016,meyer2019}; but similarly rapid variability is predicted for the X-ray regime as well \citep[e.g.,]{christie2019}.
In the context of the `mini-jets' scenario, the extremely rapid \gm-ray variability events are distinctly observed as additional flares on top of a more slowly evolving (\gm-ray) light curve.
Similarly, we would expect a coherent short-term rise and decay as explicit structures in the light curve while a mini-jet crosses our line of sight and synchrotron emission appears enhanced due to additional Doppler boosting, but we do not observe such a signature. 
Therefore, we deem such a scenario unlikely for this observed flare in Mrk\,421.
Instead of distinct features, the variability time scale and PSDs capture fluctuations in the light curve, which could either originate from $1/f^{\alpha}$ noise, or a multitude of small acceleration sites that are not resolvable in our X-ray light curve such as plasmoids in a reconnection layer in the jet that power flares when exiting the current sheet or `spine' \citep{petropoulou2016}.
When comparing our observables to the predictions, we find that an emission region with a light-travelling distance equal to the rapid variability time scale in the 0.3--10\,keV band has a size of roughly $5.7\times10^{14}$\,cm, which falls into the range of sizes for small and fast plasmoids derived by \citet{petropoulou2016}. 
While we are not able to resolve individual flares produced by these small but highly relativistic plasmoids, we might see their signatures as variability on top of the slowly increasing envelope of the flare. 
Plasmoid-dominated reconnection is also able to explain the full X-ray flare that we observe over the duration of $>15$\,hours when considering large, or so-called `monster' plasmoids \citep{christie2019}.
In addition, the merging of two plasmoids can be seen as an instant injection of particles that is biased towards high energies, which would result in, e.g., spectral hardening \citep{christie2019}, which we clearly observe in the X-ray spectra presented in this work.
We conclude that a scenario based on plasmoids undergoing magnetic reconnection in spine layer embedded in the jet is able to explain the observed X-ray flare in Mrk\,421.

The observation presented in this paper reveals a strong correlation between different X-ray bands consistent with zero lag as shown in the zDCF results. The hysteresis curve shows a clockwise rotation, which implies the electron cooling being dominant over the electron acceleration. Indeed, this energy dependence, also visible in the higher fractional variability amplitude at higher energies, implies that the rapid variability likely originates from acceleration/cooling mechanisms instead of a change in the magnetic field or the Doppler factor.
Recent X-ray polarization observations of Mrk\,421  with the Imaging X-ray Polarimetry Explorer (IXPE) yield a detailed look into the jet region where X-rays are produced. During the first observation of the source with IXPE in May 2022, the polarization degree and angle stayed constant while the flux varied, which points towards shock acceleration being the dominant mechanism responsible for the X-ray emission \citep{digesu2022}. However, two observations in June 2022 revealed a significant rotation of the X-ray polarization angle in combination with a harder-when-brighter behaviour, while no angle rotations were observed at optical or radio frequencies \citep{digesu2023}. From their observations, \citet{digesu2023} conclude that the X-ray emission site is a shock region moving along a helical magnetic field of the jet, similar to a spine-sheath scenario.

This X-ray observation of Mrk\,421 was taken during an enhanced state of VHE \gm-ray activity. While the \gm-ray data lacks the time resolution obtained with \xmm, the \gm-ray flux measured at the beginning of the \xmm observation and again shortly after, reveals in an increase in amplitude by a factor of 2 \citep{gokus2021_mrk421}.
In 2019, no instrument existed to take X-ray polarization measurements, hence we cannot know what the polarization properties during our \xmm observation were. Rapid polarization swings can be associated with \gm-ray flares \citep{zhang2022}, however given the existing FACT data, we observe, if any, only a moderate flare at VHE \gm-rays compared to past flaring activity. 

While this \xmm observation provides temporally very high-resolution X-ray data, no comparable data for the VHE \gm-ray energy range exists for the duration of this flare, hence it is not clear if Mrk\,421 exhibited variability on similar time scales at TeV energies as well. A simultaneous coverage of X-ray and TeV energies holds the potential of detecting contemporaneous rapid variability, such as predicted by plasmoid-dominated reconnection \citep{christie2019}. Testing these particle acceleration models will be feasible via multi-wavelength campaigns with the Cherenkov Telescope Array and existing as well as future X-ray missions, such as the High-Energy X-ray Probe \citep[HEX-P; ][]{hexp_white_paper}, which is designed to provide very high resolution timing data across a broad X-ray energy range from 0.2 to $\sim$150\,keV, and therefore would be ideally suited for high-energy timing studies of blazars \citep{hexp_blazars}.

\section*{Acknowledgements}
We want to express gratitude to the \xmm operations team for making it possible to change the observing mode on-the-fly in order to continue observing Mrk\,421 during this very bright X-ray flare. 
We thank the anonymous journal referee for constructive feedback that helped us improve the manuscript.
AG is thankful for valuable feedback and discussion with Manel Errando and Jim Buckley.
AG acknowledges funding by the Bundesministerium für Wirtschaft und Technologie under Deutsches Zentrum für Luft- und Raumfahrt (DLR grant number 50OR1607O) and by the German Science Foundation (DFG grant number KR 3338/4-1).
AM acknowledges partial support from Narodowe Centrum Nauki (NCN) grants 2016/23/B/ST9/03123, 2018/31/G/ST9/03224, and 2019/35/B/ST9/03944. BS acknowledges funding by the Bundesministerium für Wirtschaft und Technologie under Deutsches Zentrum für Luft- und Raumfahrt (DLR grant number 50OR2107).
The material is based upon work supported by NASA under award number 80GSFC21M0002 (CRESST II).
We made use of a collection of ISIS functions (ISISscripts) provided by ECAP/Remeis observatory and MIT (https://www.sternwarte.uni-erlangen.de/isis/).
The FACT collaboration acknowledges: The important contributions from ETH Zurich grants ETH-10.08-2 and ETH-27.12-1 as well as the funding by the Swiss SNF and the German BMBF (Verbundforschung Astro- und Astroteilchenphysik) and HAP (Helmoltz Alliance for Astroparticle Physics) to the FACT project are gratefully acknowledged. Part of this work is supported by Deutsche Forschungsgemeinschaft (DFG) within the Collaborative Research Center SFB 876 "Providing Information by Resource-Constrained Analysis", project C3. We are thankful for the very valuable contributions from E. Lorenz, D. Renker and G. Viertel
during the early phase of the FACT project. We thank the Instituto de Astrofísica de Canarias for allowing us to operate the telescope at the Observatorio del Roque de los Muchachos in La Palma, the Max-Planck-Institut für Physik for providing us with the mount of the former HEGRA CT3 telescope, and the MAGIC collaboration for their support.

\section*{Data Availability}
The data used for this work is publically available in the XMM-Newton Science Archive managed by ESA (https://nxsa.esac.esa.int/).
Extracted light curves and spectral fit results will be shared on reasonable request with the corresponding author.



\bibliographystyle{mnras}
\bibliography{mrk421}






\bsp	
\label{lastpage}
\end{document}